\documentstyle[12pt,svoss2,psfig]{book}

\begin{document}

\title{Massive Variability Surveys from Venezuela}

\author{C\'esar Brice\~no }

\affil{Centro de Investigaciones de Astronom{\'\i}a (CIDA), Apartado Postal
264, M\'erida 5101-A, Venezuela}

\begin{abstract}

   At the Venezuela National Astronomical Observatory we are carrying out 
   variability surveys spanning many hundreds of square degrees near the 
   celestial equator, using an 8k x 8k CCD Mosaic Camera optimized for
   drift-scanning on a 1m Schmidt telescope. \\
   Among the initial efforts was a project to obtain the first
   moderately deep, homogeneous sample of young stars over an area of 
   $\sim 180$sqr.deg. encompassing the entire Orion OB1 association, 
   one of the nearest and most active regions of star formation. 
   The results show that variability is a powerful technique to identify
   pre-main sequence populations, specially in sparse areas devoid of gas
   and dust.
   We are currently developing a massive database, equipped with web-based 
   data mining tools, that will make our data and results available to the 
   astronomical community.

\end{abstract}


\section{Introduction}

Recent surveys focusing on repeated observations of 
large expanses of the night sky
(e.g. ROTSE, Akerlof et al. 2000) are transforming
the study of variability phenomena 
one of the forefront research areas in modern day astrophysics.
These searches have also demonstrated
the unique value of small telescopes for this kind of research; 
in particular, the wide field Schmidt design, 
when equipped with mosaic CCD detectors, is a 
potentially ideal combination to collect multi-epoch data over 
extended areas of the sky in an efficient way. 
In this contribution I describe one of the variability surveys,
spanning many hundreds of square degrees near the
celestial equator, that are being carried at the Venezuela National
Astronomical Observatory using an 8k x 8k CCD Mosaic Camera, optimized for
drift-scanning, installed on a 1m Schmidt telescope.

\section{A Variability Survey of the Orion OB1 Association}

The study of how stars and planetary systems form and evolve
is an area in which variability surveys are particularly useful
as a means to search for pre-main sequence objects.
Young (1-10 Myr) low mass ($\la 2 \> M_{\odot}$) stars
are intrinsically variable at both optical (Herbst 1986) and near IR
wavelengths (Carpenter, Hillenbrand \& Strutskie 2001) by up to several
magnitudes, thus, variability should help to single out these objects 
among the general field population. However, so far variability has
been used mainly in follow-up studies rather than as a survey technique, 
to  collect time-series data of young objects that were
detected by some other method, like color-magnitude diagrams, 
X-ray and H$\alpha$ surveys. \\
In collaboration with Nuria Calvet (CfA), A.K. Vivas (CIDA) and 
Lee Hartmann (CfA), we have used the 8k $\times$ 8k CCD Mosaic Camera on the
Venezuela 1m Schmidt telescope (see Baltay et al. 2002 for a 
description of the camera; also Vivas, A.K. in this volume)
to conduct a large scale, 
systematic optical VRIH$\alpha$ survey spanning $\sim 180$ sqr.deg.
in the Orion OB1 Association (Fig.1), one
of the nearest ($d \sim 400$ pc, Brown et al. 1994) 
star forming regions.
Though Orion has been extensively studied in the past, existing 
optical/IR studies in Orion have concentrated mostly
on small regions such as the Orion Nebula Cluster
(ONC; e.g., Hillenbrand 1997)
and the surroundings of the star $\sigma$ Ori (Walter et al. 1998).
A few large scale studies (Wiramihardja et al. 1991; Alcal\'a et al. 1996)
have been done but did not find the more spread out, 
slightly older lower mass young stars.
Only by identifying and studying somewhat older ($\sim 10$ Myr), low-mass
stars in widely spread stellar populations, can we address
fundamental questions like the duration of the planet formation phase in
dusty disks around young stars, and the lifetimes of molecular clouds. 

\begin{figure}[ht]
\plotfiddle{Bricenoc.fig1.eps}{4in}{0}{50}{50}{-160}{-20}
\vskip -0.8in
\caption{Composite Digital Sky Survey image of Orion showing 
the total survey area of $180\Box^\circ$. The first strip we
observed, going over the
three Orion belt stars is indicated by dashed lines.
The Orion Nebula Cluster (ONC) and the
and the nebulosity around the young region NGC 2024 are indicated.}
\label{fig-1}
\end{figure}

We have developed tools for identifying variable stars using
differential photometry. With a $\chi^2$ test
and assuming a Gaussian distribution for the errors, we
consider variable only those objects for which the probability  that
the observed distribution is a result of the random errors is
$< 0.01$\%. 
The value of variability for picking out young stars is shown in Fig.2.
Follow up spectra of a large number of candidate
young stars indicate that our selection
technique is very efficient: roughly 50-60\% of our candidates
turn out to be young, low mass stars. 
Our initial results have been described in Brice\~no et al. (2001).

\begin{figure}[ht]
\plotfiddle{Bricenoc.fig2a.eps}{2.0in}{0}{30.0}{30.0}{-180.0}{-45.0}
\plotfiddle{Bricenoc.fig2b.eps}{2.0in}{0}{30.0}{30.0}{10.0}{110.0}
\plotfiddle{Bricenoc.fig2c.eps}{2.0in}{0}{50.0}{30.0}{-150.0}{100.0}
\vskip -2.2 true in
\caption{
V vs. V-I$_C$ diagram for different samples.
(a) Left: all {\bf variables} in Ori 1b;
(b) Right: all variables in a control field off the Orion region.
Solid line: ZAMS.
PMS variables in 1b are clearly separated from the background,
but they do not show in the control field.
(c) Center: coverage in RA of the variability survey centered at DEC=-3$^o$
}
\label{fig-2}
\end{figure}
\section{The Equatorial Variability Survey}

The exciting results in Orion, and those presented in this same conference by
A.K. Vivas, have prompted us to extend our spatial coverage to span all the sky
near the Celestial Equator, from 0-24h. 
We have now accumulated 4 years of data for a strip centered at $\delta=-1^o$.
Presently we are scanning a new strip centered at $\delta=-3^o$, which
we plan to observe repeatedly during the next 2 yrs. The actual 
temporal sampling
is dictated by the number of nights available for our project, usually
8-10 nights/month, and the weather (Fig. 2c).
Using the YALO telescope we have obtained data of selected fields 
along our survey strips for the photometric calibration of
the data.
The variability database will provide a unique research tool for our
studies of nearby star forming regions, RR Lyrae in
the Halo of the Milky Way, and Quasar variability, among others.
In collaboration with E.A. Ponsot at Universidad de los Andes, 
M\'erida, Venezuela, we are developing a variability database 
from the data collected in our surveys,
together with web-based tools to mine the data. 
The variability database will provide a unique research tool for our
studies of nearby star forming regions, RR Lyrae in
the Halo of the Milky Way, and Quasar variability, among others.
We expect to release an initial version of the database to
the astronomical community in the near future.

\acknowledgments
      This work is supported
by grants S1-2001001144 of FONACYT, Venezuela, AST 9987367 of the National
Science Foundation, and NASA Origins NGC5-10545, USA.

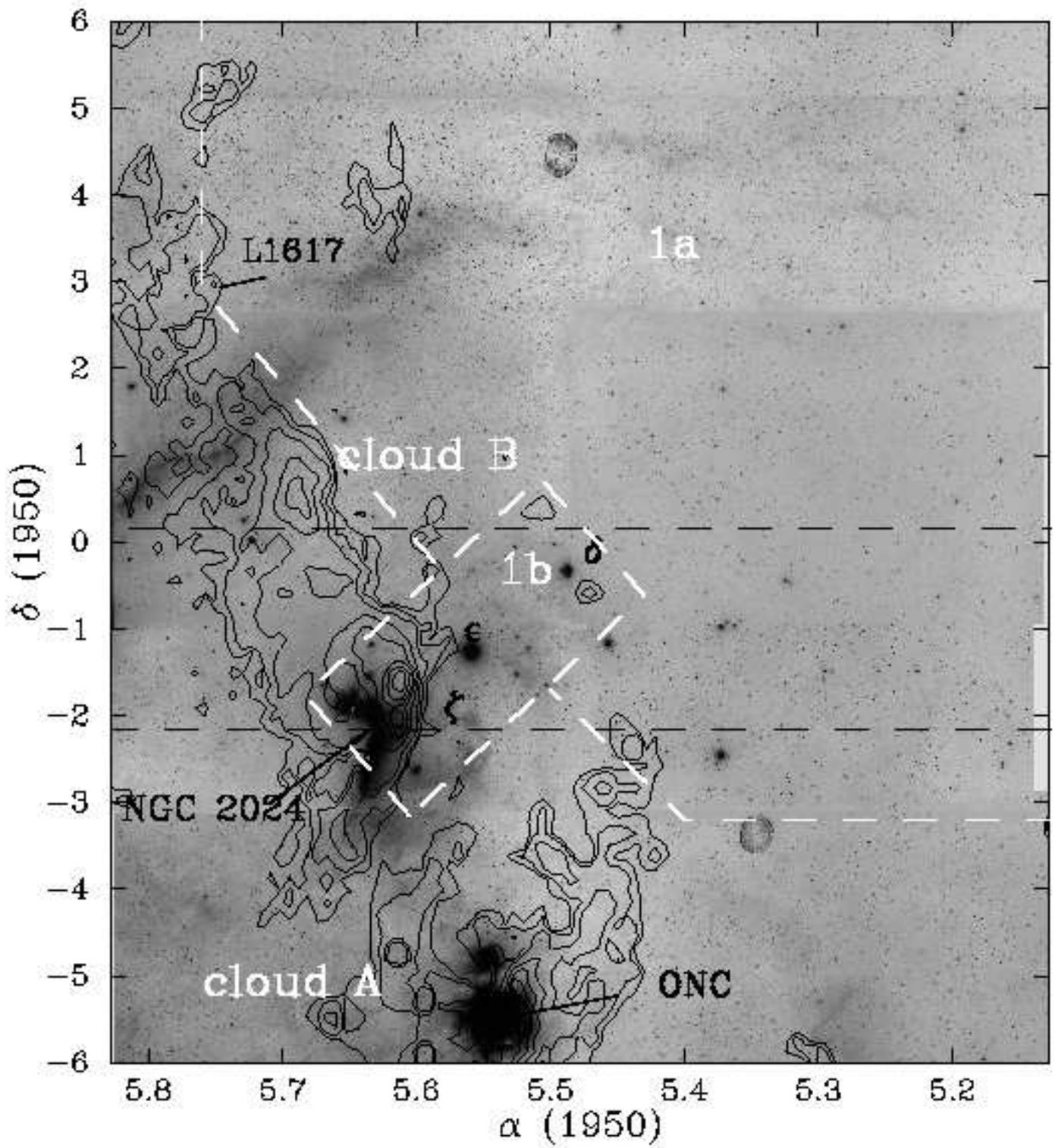
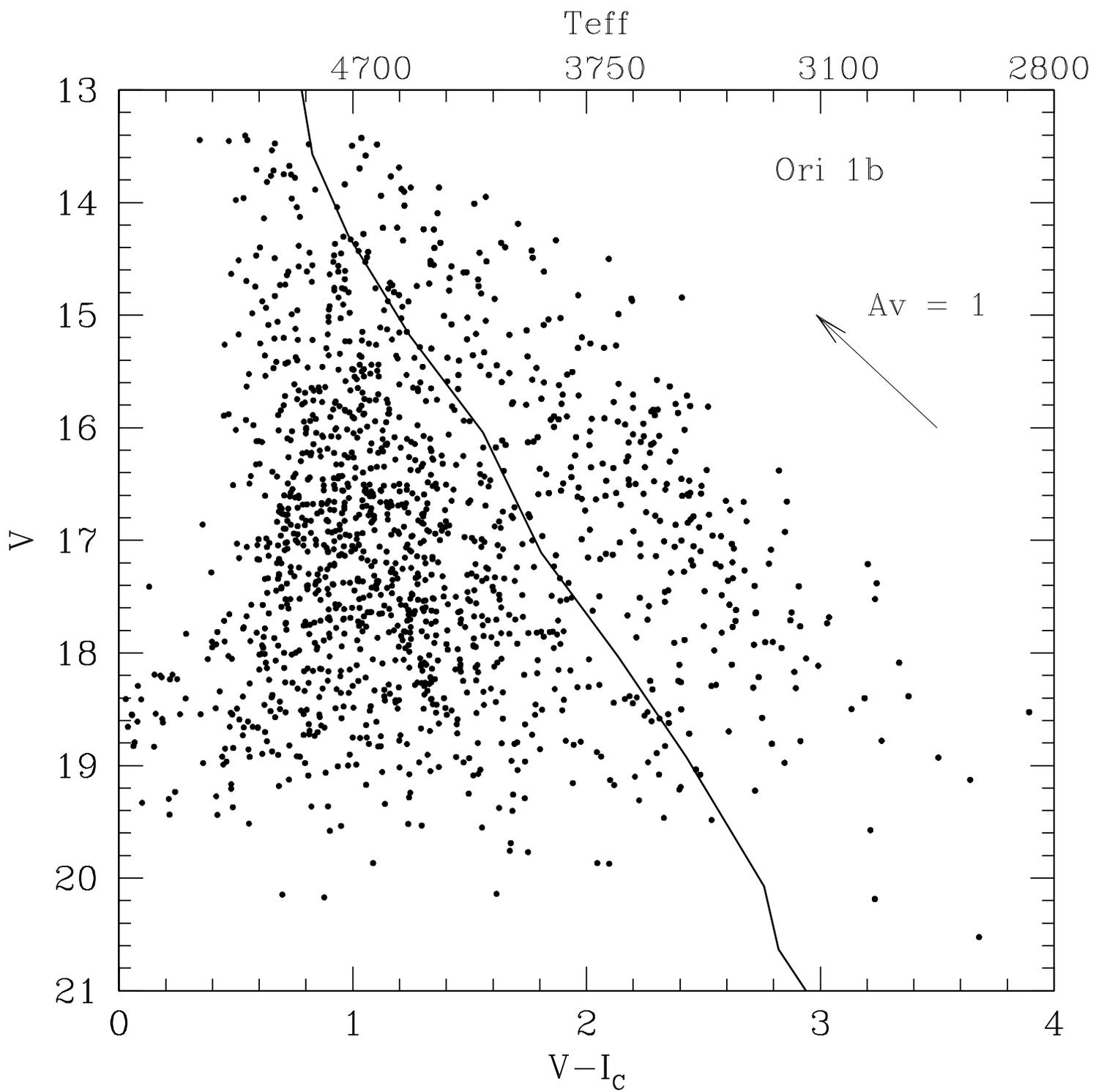
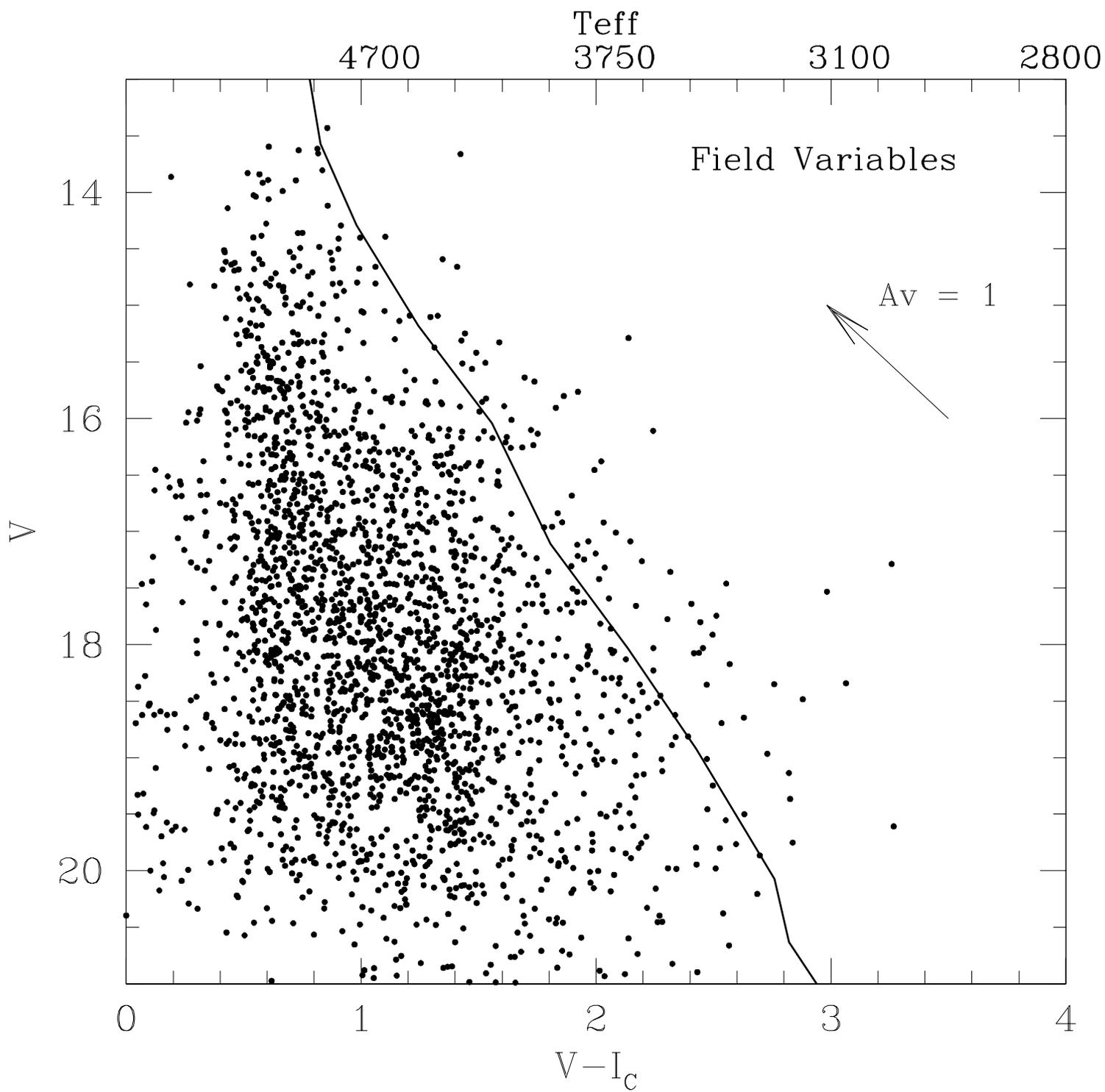
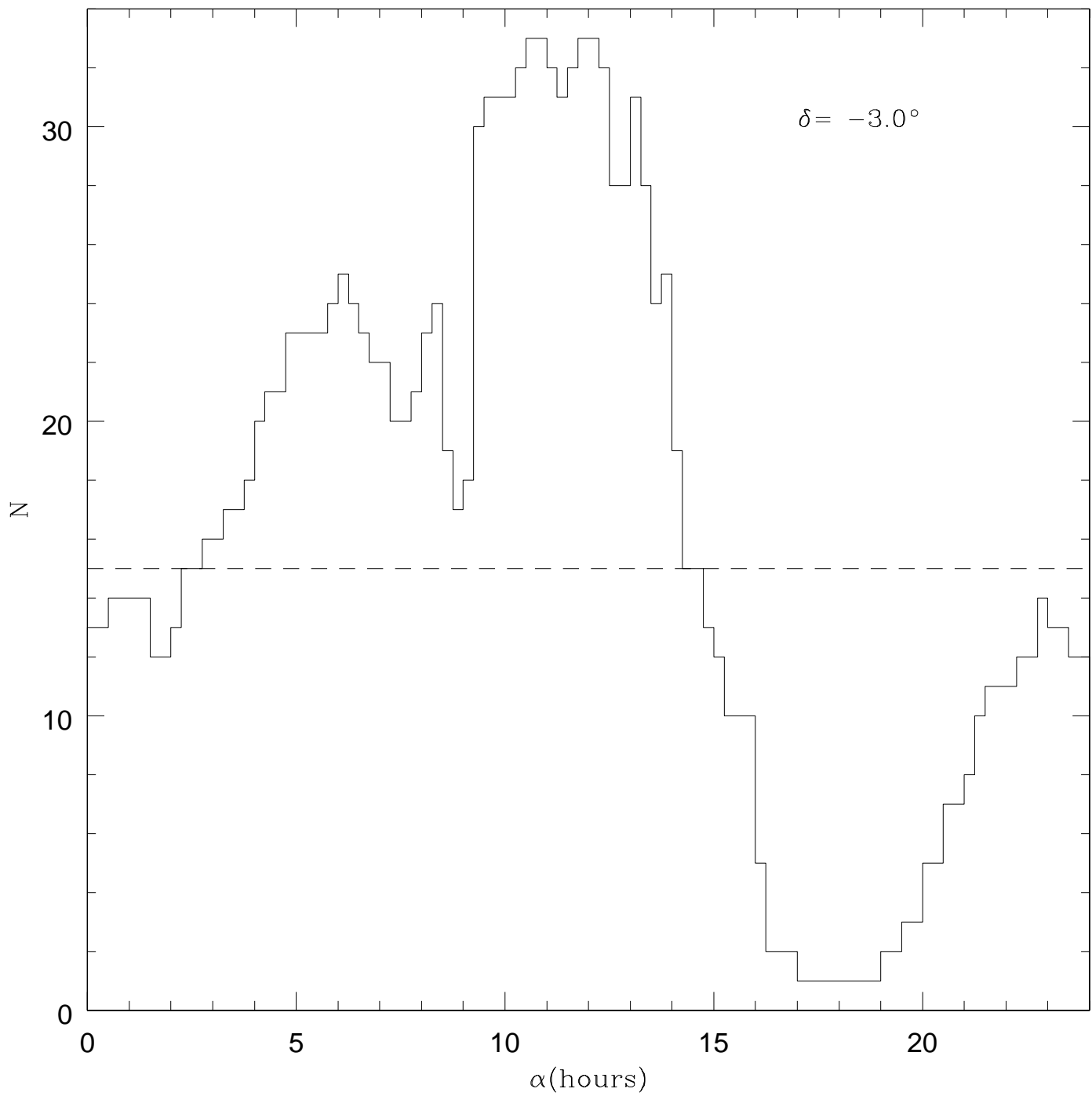
\end{document}